\documentclass{article}
\begin{document}

\begin{center}

{\Large\bf Supergravity interacting with superbranes} 

{\large\bf and spacetime
Higgs effect in General Relativity}
\footnote{To appear in ``{\it Symmetries in gravity and field theory}. 
  Conferencia homenaje en el 60 cumplea\~nos
       de Jos\'e Adolfo de Azc\'arraga.
        Salamanca. June 9-11, 2003''.} 

\medskip

{\large Igor A. Bandos }

\medskip

{\small\it
Departamento de F\'{\i}sica Te\'orica and IFIC,
 46100-Burjassot (Valencia), Spain  and \\
Institute for Theoretical Physics, NSC KIPT,
UA61108, Kharkov, Ukraine }

\end{center}

\def\theequation{\arabic{equation}}

\begin{abstract}
{

In this contribution we will review briefly the supersymmetric 
Lagrangian approach to the supergravity--superbrane interaction which
was developed in collaboration with J. A. de Azc\'arraga,
J.M. Izquierdo, J. Lukierski and J. M. Isidro. The
main accent will be made on the pure gauge nature of the 
(super)brane coordinate functions in the presence of dynamical 
(super)gravity  
described by an action rather than as a fixed background. 
This pure gauge nature just reflects the fact that the coordinate functions 
are Goldstone fields corresponding to the spontaneously broken 
diffeomorphism gauge symmetry of the interacting system. Moreover, 
a brane does not carry any local degrees of freedom in such an interacting 
system. This fact related with fundamental properties of General Relativity 
(discussed already at 1916) can be treated as 
a peculiarity of the spacetime Higgs effect which occurs 
in General Relativity in the presence of  material particles, 
strings and branes. 
}

\end{abstract}

{\bf 1. Introduction}.
Let me begin by thanking the organizers for the nice opportunity to speak  
at this conference dedicated to Jos\'e Adolfo de Azcarraga. I had already 
known him through his papers for more than 
20 years. At that time, working in Kharkov in the group of Dmitri Volkov, we 
had been especially influenced by the work \cite{AL}, where 
$\kappa$--symmetry was found, and later by the topological treatment of
tensorial 'central' charges of the most general supersymmetry algebra
\cite{AT} (see the contribution of D. Sorokin  \cite{DS} 
for more details). 
In the last few years I had the pleasure to know Jos\'e Adolfo personally, 
to enjoy his friendship and to collaborate with him. I have enjoyed very much 
discussing with Jos\'e, 
and not only on physics:  he has widespread interests in many 
other areas, both scientific and non--scientific.

Our collaboration has been not only pleasant but quite productive
\cite{BPS01,BdAIL1,BdAI,BdAIL2,BdAIL2',BdAIL3}.
Interestingly enough, the field of our studies can be separated into 
two streams, one \cite{BPS01} being related to the work \cite{AT} 
(as well as with more recent studies of \cite{AI02}) 
and the other
\cite{BdAIL1,BdAI,BdAIL2,BdAIL2',BdAIL3}, dealing with 
the Lagrangian description of dynamical supergravity interacting with 
supersymmetric extended objects,  is related
to the $\kappa$--symmetry of superbrane (see below), and thus,
with the work of \cite{AL}.  This latter direction 
will be the main subject of the present contribution. 

The results of  \cite{BdAI,BdAIL2,BdAIL2'} 
have a clear `bosonic projection' which
appeared to be related \cite{BdAIL3} 
with fundamental properties of General
Relativity (quite well known and actively discussed decades ago
\cite{Einstein,EiGro,Lusanna}).
In the modern language, they may be treated as an analysis of 
a spacetime counterpart of Higgs effect which occurs 
in general relativity in the presence of material particles, strings  
or $p$--branes.

\bigskip

{\bf 2. Problem of supersymmetric Lagrangian description of
super\-gravi\-ty--superbrane interaction.}
A popular description of superbranes was proposed in 1989
\cite{DGHR,SUGRA}.
It identified them with solitonic
solutions of the pure bosonic `limit' of the supergravity equations.
Although all the fermions are set equal to zero, the superbrane solutions are
supersymmetric and, hence, stable.

On the other hand, one may also consider a
superstring or superbrane in a curved superspace defined by a
supergravity {\sl background} \cite{strKsg,BST87}. Then,
selfconsistency requires having a smooth flat superspace limit for
such a system. This implies, in particular, that the superbrane
has to  possess local fermionic $\kappa$--symmetry
\cite{AL,Siegel} in curved superspace, as it does in the flat one. 
Such a requirement immediately results in 
superfield supergravity constraints being imposed on the
background superfields \cite{strKsg,BST87}. The point is that, for 
the most interesting (in M-theoretical perspective) $D=10,11$ cases these are
the {\sl on-shell} constraints: their selfconsistency
implies equations of motion for $D=10$ supergravity, 
and these are sourceless or `free' equations. Clearly, such a description 
is an approximate one.  

In purely bosonic `limit' one can describe 
the interacting system of dynamical gravity and a $p$--dimensional 
material object ($p$--brane) by the action 
\begin{eqnarray}
\label{SEH+Sp}
S &=& S_{EH\, D} +  S_{p\, D}
= {1\over 2\kappa}\int d^D x \, \sqrt{|g|} {\cal R} + S_{p\, D} \; ,
\\ \label{SpD}
  S_{p\, D} &=& {T_p\over 4}\int d^{p+1} \xi
[\sqrt{|\gamma|} 
\gamma^{mn}(\xi)
\partial_m\hat{x}^\mu \partial_n \hat{x}^\nu
g_{\mu\nu}(\hat{x}) + (p-1)\sqrt{|\gamma|} ] \quad \; .
\end{eqnarray}
which is the sum of Einstein--Hilbert action for gravity $S_{EH\, D}$ 
and the bosonic $p$--brane action $S_{p\, D}$ written in terms of 
coordinate functions $\hat{x}^\nu(\xi)$ determining the position of the 
$p$--brane 
worldvolume $W^{p+1}$ in the spacetime [$\xi^m=(\tau , \sigma^1, \ldots 
\sigma^p)$ are local coordinates on  $W^{p+1}$, 
$\gamma_{mn}(\xi)$ is an auxiliary worldvolume metric; on the mass 
shell it coinsides with the induced metric, 
$\gamma_{mn}(\xi)=\partial_m \hat{x}^\mu \partial_n \hat{x}^\nu 
g_{\mu\nu}(\hat{x})$]. 
The variation of the action 
with respect to the metric $g_{\mu\nu}(x)$ 
(see \cite{SUGRA}) 
produces the Einstein equation 
\begin{eqnarray}\label{EiEq0}
{\cal G}_{\mu\nu}\equiv \sqrt{|g|}
({\cal R}_{\mu\nu}- {1\over 2} g_{\mu\nu}{\cal R}) =
\kappa \,T_{\mu\nu} \;\; 
\end{eqnarray}
with the singular 
energy--momentum tensor from the $p$--brane 
\begin{eqnarray}\label{Tmnp}
& T^{\mu \nu} = {T_p \over 4} 
\int d^{p+1}\xi \sqrt{\gamma} \gamma^{mn}
\partial_m{\hat{x}}^\mu \partial_n{\hat{x}}^\nu  
\delta^{D} (x-\hat{x}(\xi))   \; ,
 \end{eqnarray} 
while the variation with respect to 
coordinate functions $\hat{x}^\mu(\xi)$ produces 
the equations of motion for the $p$--brane 
\begin{eqnarray}\label{pgeod}
\partial_{m} (\sqrt{|\gamma|}\gamma^{mn}g_{\mu\nu}(\hat{x}) 
\partial_{n}{\hat{x}}^\nu (\xi))   -{1\over 2} \sqrt{|\gamma|}\gamma^{mn} 
\partial_{m}{\hat{x}}^\nu  
\partial_{n}{\hat{x}}^\rho (\partial_{\mu} g_{\nu\rho})(\hat{x})=0 \; .
\quad  
 \end{eqnarray}

Is it possible 
to provide a full (quasi)classical Lagrangian description 
of the supergravity--superbrane interacting system
similar to the description of the gravity--bosonic brane interaction given by 
the action (\ref{SEH+Sp})? 
It was natural to assume that it  is based on the sum 
$$ S= S_{SG} + S^s_p$$
of 
{\sl some} supergravity action
$S_{SG}$ and the super--$p$--brane action $S^s_p$, the same as 
in the case of supergravity background.

A superbrane is a brane moving in superspace
$Z^M = (x^\mu , \theta^{\check{\alpha}})$, {\it i.e.} 
its worldvolume ${\cal W}^{p+1}$ is a surface
in superspace which can be defined by a set of bosonic and fermionic 
coordinate functions $\hat{Z}^M (\xi^m)= (\hat{x}^\mu (\xi),
 \hat{\theta}^{\check{\alpha}} (\xi))$, 
\begin{eqnarray}
  \label{Wp+1}
{\cal W}^{p+1} \; \subset \; \Sigma^{(D|n)}\; : \
Z^M= \hat{Z}^M (\xi^m):= (\hat{x}^\mu (\xi),
 \hat{\theta}^{\check{\alpha}} (\xi))
\; ,   \quad m=0, \ldots, p\, ,
\end{eqnarray}
The action $S_p^s$ of a super--$p$--brane in 
a curved superspace  \cite{strKsg,BST87,AETW} is formulated in terms 
of supergravity superfields, 
supervielbeins 
$E^A (Z)= dZ^M E_M^A (Z)=(E^a, E^\alpha )$ 
and superform gauge superfields $C_q(Z) = {1\over q!} dZ^{M_q} \wedge
\ldots \wedge dZ^{M_1} C_{M_1\ldots M_q}(Z)$. 
For instance, 
for the super--$p$--branes of the 
`old brane scan' \cite{AETW} one has  
$ S^s_p = 1/4 *\hat{E}_a\wedge \hat{E}^a - (-)^p(p-1)/4 *1 -\hat{C}_{p+1}$, 
or, equivalently, 
 \begin{eqnarray}
  \label{Sps}
S^s_p & = \int\limits^{}_{{\cal W}^{p+1}} 
d^{p+1} \xi
\sqrt{|\gamma|} \left[\gamma^{mn}(\xi)
\partial_n\hat{Z}^N \partial_m \hat{Z}^M {E}_{M}^a(\hat{Z})
{E}_{N}^b(\hat{Z})\eta_{ab}    
+ (p-1)\right] - \quad \nonumber \\  
& - \int\limits^{}_{{\cal W}^{p+1}} {1\over p!} 
\varepsilon^{m_1\ldots m_{p+1}} \partial_{m_{p+1}} \hat{Z}^{M_{p+1}} 
\ldots  \partial_{m_1} \hat{Z}^{M_1} C_{M_1\ldots M_{p+1}}(\hat{Z}) \; . 
\end{eqnarray}
Note that the superbrane action 
involves {\sl only bosonic superforms};
the fermionic supervielbein $E^\alpha= dZ^M E_M^\alpha(Z)$ is not present
in $S_{p}= S_p[\hat{E}^a, \hat{C}_{p+1}]$. 
Here and below the hats over superfields or superforms  
indicate dependence on the coordinate functions {\it i.e.},  
$\hat{Z}^M (\xi)= (\hat{x}^\mu (\xi),  
 \hat{\theta}^{\check{\alpha}} (\xi))$ replace   
the superspace coordinates
$Z^M = (x^\mu , \theta^{\check{\alpha}})$. The hat also distinguishes  
between the coordinate functions and the spacetime or superspace coordinates.

To be able to derive the equation of motion, 
one should have supergravity part $S_{SG}$ of the action 
$S=S_{SG}+S_p[\hat{E}^a, \hat{C}_q]$  
formulated in terms of the same variables; {\it i.e.} one should 
use  the {\sl superfield supergravity action} $S_{SG}[E^A, C_q]$, 
 \begin{eqnarray}
  \label{S=sf}
S= S_{SG}[E^A, C_q]  + S_{p}[\hat{E}^a, \hat{C}_q] \equiv  
S_{SG}[E^A(Z), C_q(Z)]  + S^s_{p}[{E}^a(\hat{Z}),
{C}_q(\hat{Z}] \; , \qquad 
\end{eqnarray}

In the $D=4$, $N=1$ case the superfield action for supergravity is known
[$S_{SG}[\hat{E}^A] =\int d^8 Z sdet(E_M^A(Z))$] and one can study  
the interacting  system (\ref{S=sf}).
This has been done for supergravity--superparticle 
and supergravity---superstring
interacting systems \cite{BdAIL2,BI03}.
The problem, however, is that for the most interesting 
$D=10$ and $D=11$ cases  no {\sl superfield} action for supergravity is
known.
\footnote{As, in contrast, the {\sl component} action $S_{SG}[e^a,
dx^\mu \psi_\mu^{\alpha}, \ldots]$ for all $D\leq 11$
supergravity theories  is now known, one might think of using it
in (\ref{S=sf}) instead of the superfield supergravity action,
and decompose the superfields in
 $S_{p}[\hat{E}^a, \hat{C}_q]$ in terms of component fields, {\it e.g.}
$E^a_\mu= e^a_\mu(x) + {\cal O}(\theta )$, $E_\mu^{\alpha}=
\psi_\mu^{\alpha}(x)+ {\cal O}(\theta) $. However, despite the
first impression the va\-ri\-ational problem for the action
$S_{SG}[e^a, dx^\mu \psi_\mu^{\alpha},\ldots ] + S_{p}[\hat{E}^a,
\hat{C}_q]$ is not well posed, see \cite{BdAIL2'}.}

\bigskip

{\bf 3. A gauge--fixed description of dynamical supergravity interacting with
material superparticles, superstrings and superbranes}

{\it 3.1. A practical way to study the supergravity--superbrane interacting
system} without a use of the superfield supergravity action was
proposed in \cite{BdAIL2,BdAIL2'}. 
It was shown there that the dynamical
system described by the superfield action (\ref{S=sf}) (still
hypothetical and, possibly, non existing in $D=10, 11$) is gauge
equivalent to the more simple dynamical system described by the sum
\begin{equation}\label{SG+bb}
S= S_{SG}[e^a(x), dx^\mu \psi_\mu^\alpha(x) , C_q(x)] +
S_p[\hat{e}^a, \hat{C}_{q}] \;  ,
\end{equation}
of the spacetime  (component) action for supergravity
$S_{SG}[e^a(x), \psi^\alpha(x) , C_q(x)]$ {\it without auxiliary
fields} and the action for the bosonic brane, $S_p[\hat{e}^a,
\hat{C}_{q}(\hat{x})]$, which is given by a purely bosonic 'limit'
of the superbrane action $S_p[\hat{e}^a, \hat{C}_{q}]\equiv\;\;$ 
$S^s_p[\hat{E}^a,\hat{C}_{q}(\hat{Z})]\vert_{\hat{\theta}=0}$. Such
a gauge fixed description is complete in the sense that it
produces the gauge fixed version of all the dynamical equations of
the complete interacting system \cite{BdAI,BdAIL2}. It also
possesses the $1/2$ of the local supersymmetry characteristic of
the 'free' supergravity action $S_{SG}[e^a,\psi^\alpha, C_q(x)]$
\cite{BdAI}. This preserved $1/2$ of the supersymmetry reflects
the $\kappa$--symmetry of the prototype  superbrane action
\cite{BdAI}.

\medskip

{\it 3.2. Pure gauge nature of the superbrane coordinate functions 
in the presence of dynamical supergravity.} 
What is stated above implies that the fermionic {\it coordinate functions}
of a superbrane,  $\hat{\theta}^{\check{\alpha}} (\xi)$ (not to be confused
with the superspace coordinates ${\theta}^{\check{\alpha}}$), 
{\it have a pure gauge
nature}. The gauge symmetry which can be used to gauge it away
\begin{equation}
  \label{thGAUGE}
\hat{\theta}^{\check{\alpha}}(\xi)= 0\; .
\end{equation}
is the super{\it diffeomorpism} symmetry
or {\it passive form of the} superspace
{\it general coordinate invariance}. This implies arbitrary, but invertible,
transformations of the superspace coordinates
\begin{eqnarray}
  \label{sdif}
\delta_{sdiff} x^\mu &=& b^\mu (x, {\theta})= b^\mu(x) + {\cal O}(\theta)
\; , \qquad
\\
\delta_{sdiff} {\theta}^{\check{\alpha}} &=&
b^{\check{\alpha}}(x, {\theta})
= \varepsilon^{\check{\alpha}}(x) + {\cal O}({\theta}) \; ,
\end{eqnarray}
but leaves invariant differential forms,
{\it e.g.} $E^{A \prime} (Z^\prime) = E^A(Z)$, where 
$Z^{M\prime}= Z^M + \delta_{sdiff}Z^M= Z^M + b^M(Z)$.
The leading terms 
$b^\mu(x)= b^\mu (x, 0)$ and $\varepsilon^{\check{\alpha}}(x)
=b^{\check{\alpha}}(x, 0)= b^{\check{\alpha}}\vert_{\theta=0}$
in the decomposition of the superdiffeomorphism parameters
$b^M(Z)=(b^\mu(Z),b^{\check{\alpha}}(Z))$
can be identified with the parameters of spacetime diffeomorphisms
($b^\mu(x)$) and of the
{\it local supersymmetry} ($\varepsilon^{\check{\alpha}}(x)$), 
 of the {\it component} (spacetime) formulation
of the supergravity. 

In the supergravity--superbrane interacting system described by the action
(\ref{S=sf}) the superdiffeomorphisms act also on the coordinate function
by $\delta_{sdiff} \hat{x}^{\mu}(\xi)=
b^{\mu}(\hat{x}(\xi), \hat{\theta}(\xi))$,
$\delta_{sdiff} \hat{\theta}^{\check{\alpha}}(\xi)=
b^{\check{\alpha}}(\hat{x}(\xi), \hat{\theta}(\xi))$.
The interacting system also possesses  {\it repara\-metri\-zation} 
symmetry 
(or worldvolume diffeomorphism invariance) [$\delta_{rep}\xi^m =
\beta^m(\xi)$, $\hat{Z}^{\prime M}(\xi + \delta_{rep}\xi) = \hat{Z}^M(\xi)$]  
 and {\it $\kappa$--symmetry}
[$\delta_{\kappa}\hat{Z}^M= \kappa^\alpha (1-\bar{\Gamma})_\alpha{}^\beta
E_\beta{}^M(\hat{Z})$ with a certain 
projector $(1-\bar{\Gamma})_\alpha{}^\beta$
($tr \bar{\Gamma} =0$, $\bar{\Gamma}\bar{\Gamma}=I$)], 
both acting on the coordinate functions only. Thus the complete set of gauge
transformations acting on the coordinate functions read
\begin{eqnarray}
\label{sdifB+k}
\delta_{gauge} \hat{x}^{\mu}(\xi)=
b^{\mu}(\hat{x}(\xi)), \hat{\theta}(\xi))+
\delta_{\kappa}\hat{x}^{\mu}(\xi)  + \delta_{rep}\hat{x}^{\mu}(\xi)
\\ \label{sdifF+k}
\delta_{gauge} \hat{\theta}^{\check{\alpha}}(\xi)=
b^{\check{\alpha}}(\hat{x}(\xi)), \hat{\theta}(\xi)) +
\delta_{\kappa}\hat{\theta}^{\check{\alpha}}(\xi)+
\delta_{rep}
\hat{\theta}^{\check{\alpha}}(\xi)\; .
\end{eqnarray}

The gauge (\ref{thGAUGE}) can be fixed using the fermionic
superdiffeomorphisms (actually its leading component, the local supersymmetry
$\varepsilon^{\check{\alpha}}(x)$). Then the symmetry preserving the
gauge (\ref{thGAUGE})  is defined by
$\delta_{gauge} \hat{\theta}^{\check{\alpha}}(\xi)\vert_{\hat{\theta}=0}= 0$,
which (in the light of $\delta_{rep} \hat{\theta}^{\check{\alpha}}(\xi)
\vert_{\hat{\theta}=0}\equiv 0$) implies
\begin{eqnarray}
\label{thGcons}
\delta_{gauge} \hat{\theta}^{\check{\alpha}}(\xi)\vert_{\hat{\theta}=0}= 0
\quad \Rightarrow \qquad
\varepsilon^{\check{\alpha}}(\hat{x}) =
-\delta_{\kappa}\hat{\theta}^{\check{\alpha}}(\xi)\vert_{\hat{\theta}=0}
\equiv - \kappa^\alpha (1-\bar{\gamma})_\alpha{}^\beta
\delta_\beta{}^{\check{\alpha}}\; .
\end{eqnarray}
This shows the preservation of
the $1/2$ of the local supersymmetry on the worldvolume as well
as the relation  of this preserved supersymmetry with
 the $\kappa$--symmetry of the parent superbrane action
[$\bar{\gamma}= \bar{\Gamma}\vert_{\hat{\theta}=0}$ is the
`leading component' of the parent $\kappa$--symmetry projector, 
see \cite{BdAI}].

\bigskip

{\it 3.3. How to arrive at the gauge fixed action (\ref{SG+bb}).}

An important property of the gauge (\ref{thGAUGE}) is that it
can be fixed simultaneously with the Wess--Zumino (WZ) gauge for supergravity
\footnote{See \cite{BdAIL2} and refs therein; the WZ gauge implies,
in particular, $E_\beta{}^{\check{\alpha}}(x, 0) =
\delta_\beta{}^{\check{\alpha}}$ which has been already used in the second form
of Eq. (\ref{thGcons}).}. In the WZ gauge  one can
perform the Grassmann integration 
in the superfield supergravity action $S_{SG}$ and arrive at a spacetime
(component) action for supergravity,
but with auxiliary fields,
$S_{SG}[e^a(x), \psi^\alpha(x), C_q(x), aux.fields]$.
Then one has to observe that {\it i)} the leading components of all 
superfields/superforms of the superfield supergravity
are certain physical fields of  the 
supergravity multiplet, {\it e.g.} 
$E^a\vert_{\theta=0, d\theta=0}= dx^\mu e_\mu^a(x)$, 
$E^\alpha\vert_{\theta=0, d\theta=0}= dx^\mu \psi_\mu^\alpha(x)$, 
{\it ii)} all the super--$p$--brane actions
$S^s_{p}$ (including the ones from the
`old brane scan' \cite{BST87,AETW}, Eq. (\ref{Sps}) as well as 
Dirichlet superbranes and M5--brane) 
involve only bosonic superfields of the supergravity, 
$S^s_{p}=S_p[E^a(\hat{Z}), C_q(\hat{Z})]$.
Thus in the gauge (\ref{thGAUGE}) the superbrane action $S_p^s$ 
becomes 
the action for a bosonic brane $S_{p}[e^a(\hat{x}), C_q(\hat{x})]
=S^s_{p}[E^a(\hat{Z}), C_q(\hat{Z})]\vert_{\hat{\theta}=0}$ involving
{\sl only physical bosonic fields of the supergravity multiplet}.
No auxiliary fields enter in this bosonic brane action.
As so, the auxiliary fields can be removed from 
$S_{SG}^0(e^a(x), \psi^\alpha(x), C_q(x), aux.fields)$ by using their purely
algebraic equations of motion, in the same manner as it is done in the case of
`free' supergravity.

After this stage one arrives at the gauge fixed action of the form
(\ref{SG+bb}). 

The gauge fixed description of the supergravity---superbrane
interacting system provided by this action is quite practical
because {\it i)} 
it involves only the component action for supergravity known
in all dimensions including $D=10,11$, {\it ii)} it provides a gauge fixed
description of the supergravity--superbrane interacting system which is
{\it complete} in the sense that it produces the gauge fixed versions of all
the dynamical equations following from the original superfield action
(\ref{S=sf}) \cite{BdAI,BdAIL2} [the reasons for this 
may be understood from the discussion below]; 
{\it iii)} it
still possesses $1/2$ of the local supersymmetry of the `free' supergravity
action; moreover, this `preserved' $1/2$ of the local supersymmetry, on one
hand, reminisces the $\kappa$--symmetry of the
prototype super--$p$--brane  action and, on the other hand,
is related with supersymmetry preserved by the super--$p$--brane BPS state as
described by (pure bosonic) solitonic solution of supergravity.

\bigskip

{\bf 4. The fate of the brane degrees of freedom
in General Relativity interacting with material particles,
strings and branes.}

The above consideration on the  pure gauge nature of the 
fermionic coordinate functions
of a superbrane has a clear bosonic counterpart. The bosonic
(super)diffeo\-morphisms act additively on the bosonic coordinate function
$\hat{x}^\mu(\xi)$, Eq. (\ref{sdifB+k}). Certainly, in contrast with fermionic
case, one cannot set all the coordinate functions to zero [as
diffeomorphisms are invertible transformations and, hence, cannot map
a (region of a) surface into a point\footnote{In contrast, the gauge 
$\hat{\theta}^{\check{\alpha}}(\xi) =0$ is fixed by using 
$b^{\check{\alpha}}(\hat{x}, 0)$ parameter, and this does not enter in the 
matrix 
$\partial b^{\check{\alpha}}(x,\theta)/
 \partial {\theta}^{\check{\gamma}}$
which has to be nondegenerate to provide invertibility 
of the superdiffeomorphism transformations.}].
However,  one can use the diffeomorphisms
to fix
{\sl locally} (in a tubular neighborhood of a point of ${\cal W}^{p+1}$)
the {\it static gauge} \cite{BdAIL3}
\begin{eqnarray}\label{gaugeXp}
\hat{x}^\mu &=& (\xi^m , \vec{0})= (\tau , \sigma^1 , \ldots , \sigma^p ,
0 , \ldots , 0)\; ,
\end{eqnarray}
where first $(p+1)$ coordinate functions are identified
with the local worldvolume coordinates,
$\hat{x}^m(\xi) = \xi^m$, while the remaining $(D-p-1)$ coordinate functions
vanish, $\hat{x}^I(\xi) = 0$.

All this is true also in pure bosonic case, which corresponds to
General Relativity interacting with material particles, strings or branes
described by the action  (\ref{SEH+Sp}), (\ref{SpD}). 
Thus, this property should be well known in general relativity, at least
for the $D=4$, $p=0$ case corresponding to General Relativity interacting 
with a material particle
\begin{eqnarray}\label{SEH+Sm}
& S  = S_{EH}+ S_{0\, m} :=  
S_{EH} + {1\over 2}  \int d\tau (l(\tau) g_{\mu\nu}(\hat{x})
\dot{\hat{x}}^\mu(\tau) \dot{\hat{x}}^\nu(\tau) + l^{-1}(\tau)
m^2)\; . \quad 
\end{eqnarray}
This is indeed the case  \cite{BdAIL3}:
the pure gauge nature of the particle coordinate function was known
already in 1926 \cite{EiGro}, but, clearly,
was discussed using different terminology.
It is well known that the Bianchi identity
${\cal G}^\mu_{\nu ;\mu}\equiv 0$ for the Einstein tensor
{\sl density} ${\cal G}^{\mu\nu}$ in the gravitational field equations 
(\ref{EiEq0}) implies the covariant conservation of the energy--momentum
tensor density 
\begin{eqnarray}\label{DT0=0}
& T^{\mu}{}_{\nu ; \mu}\equiv
\partial_\mu (T^{\mu\rho}g_{\rho \nu}) - {1\over 2}
T^{\mu\rho}\partial_\nu g_{\rho \mu} =0\; .
\end{eqnarray}
For a particle $T_{\mu\nu}$ has support on the worldline ${\cal W}^1$,
\begin{eqnarray}\label{Tmn}
& T^{\mu \nu}  =
{1\over 2} \int d\tau l(\tau)
\dot{\hat{x}}^\mu(\tau) \dot{\hat{x}}^\nu(\tau)  \delta^{4} (x-\hat{x}(\tau))
\; ,
 \end{eqnarray}
and then Eq. (\ref{DT0=0}) is equivalent \cite{EiGro} to the
particle (geodesic) equations 
\begin{eqnarray}\label{geod}
& {\cal S}_{\tau\, \mu}:=  \partial_{\tau} (l(\tau)g_{\mu\nu}(\hat{x})
\dot{\hat{x}}^\nu)  -{l(\tau)\over 2}\dot{\hat{x}}^\nu
\dot{\hat{x}}^\rho (\partial_{\mu} g_{\nu\rho})(\hat{x})=0
 \end{eqnarray}
or $d^2 \hat{x}^\mu/ds^2 + \Gamma^{\mu}_{\nu \rho} \,
d\hat{x}^\nu/ds \, d\hat{x}^\rho/ ds=0$ for $ds= d\tau /l(\tau)$.
The counterparts of the above statement 
for the cases of
gravity interacting with strings and $p$--branes were given in 
\cite{Gursey} and \cite{BdAI}.

This result exhibits a dependence among Eqs. (\ref{geod}) and
(\ref{EiEq0}) which, by the second Noether theorem,
reflects a gauge symmetry of the
action (\ref{SEH+Sm}) leading to 
Eqs. (\ref{EiEq0}) and (\ref{geod}) when varied with respect to the metric
$g_{\mu\nu}$ and with respect to the particle coordinate functions
$\hat{x}^\mu$, respectively. 
This gauge symmetry is just the 
{\it diffeomorphism invariance} (the freedom of choosing a
{\it local coordinate system}) or
{\it passive} form of the general coordinate invariance ({\it cf.} Eq. 
(\ref{sdif})) 
\begin{eqnarray}\label{xdiff}
\delta {x}^\mu  \equiv {x}^{\mu \prime}- {x}^\mu & =& b^\mu (x) \; ,
\\
\label{gdiff}
\delta^{\prime} g_{\mu\nu} (x)\equiv
g^{\prime}_{\mu\nu} (x) - g_{\mu\nu} (x) & = & - (b_{\mu ; \nu} +
b_{\nu ; \mu}) \nonumber \\ 
& \equiv 
& - (\partial_\mu b^\rho g_{\nu\rho}  + \partial_\nu b^\rho g_{\mu\rho}
+ b^\rho \partial_\rho g_{\mu\nu})  \;  , \quad 
\\
\label{X0diff}
 \delta \hat{x}^\mu (\tau)\equiv \hat{x}^{\mu \prime}(\tau )-
\hat{x}^\mu(\tau)& =&
b^\mu (\hat{x}(\tau)) \; .
\end{eqnarray}
Indeed, the general variation of the action (\ref{vSEH+Sm}) reads
\begin{eqnarray}\label{vSEH+Sm}
\delta S  &=
- {1\over 2\kappa} \int d^4 x {\cal G}^{\mu\nu} \delta^\prime g_{\mu\nu}(x) +
 {1\over 2} \int d^4 x T^{\mu\nu} \delta^\prime g_{\mu\nu}(x)
- \int d\tau {\cal S}_{\tau\, \mu} (\tau)
\delta {\hat{x}}^\mu(\tau) \; , \quad 
\end{eqnarray}
where ${\cal G}_{\mu\nu}$, $T_{\mu\nu}$ and $ {\cal S}_{\tau\, \mu}$
are defined in Eqs. (\ref{EiEq0}),  (\ref{Tmn}) and (\ref{geod}), respectively.
Substituting  (\ref{gdiff}), (\ref{X0diff}), one finds that 
 the first term vanishes since 
${\cal G}^{\mu}{}_{\nu ; \mu}\equiv 0$,
and that the second and third terms cancel (using
the second form of Eq. (\ref{gdiff})). 

Note, that when `proved the second Noether theorem' for
diffeomorphisms, in (\ref{vSEH+Sm}) we did not use the
the variation $\delta x^\mu$ (\ref{xdiff}), which is nonvanishing for the
diffeomorphism transformations. We are allowed to do this because
this variation, when considered 
without the additional ones (\ref{gdiff}),
(\ref{X0diff}),
\begin{eqnarray}\label{xgc}
&\delta {x}^\mu  \equiv {x}^{\mu \prime}- {x}^\mu = t^\mu (x) \; ,
\\
\label{ggc}
& \delta^{\prime} g_{\mu\nu} (x)\equiv
g^{\prime}_{\mu\nu} (x) - g_{\mu\nu} (x)= 0 \; , \qquad
\delta \hat{x}^\mu (\xi)=0 \; ,
\end{eqnarray}
also leaves invariant the Einstein action (and does
not act on the particle  action). 
This is the {\it active form} of general coordinate invariance
(see \cite{Einstein} and  \cite{BdAI,Lusanna}). 

Actually, one can easily prove (see \cite{BdAI}) that {\sl {\it any} 
action invariant under  
diffeomorphism {\it gauge} symmetry is automatically gauge invariant under
the active form of general coordinate symmetry (\ref{xgc}), 
(\ref{ggc})}. 
This fact had been known and led to a conceptual discussion 
on  the lack of physical meaning of the spacetime point notion 
in general relativity \cite{Einstein} (see also \cite{Lusanna}). 
The passive form of the general coordinate
invariance (diffeomorphism symmetry), Eqs.  (\ref{xdiff}), (\ref{gdiff}),
(\ref{X0diff}), was desirable as it provided
a realization of the general relativity principle. 
However, the fact that the Einstein--Hilbert action and the Einstein equations
are invariant as well under the {\it active} form of the general
coordinate transformations, {\it i.e. under a change of `physical'
spacetime  points}, Eqs. (\ref{xgc}), (\ref{ggc}),
{\it ``takes away from space and
time the last remnant of physical objectivity''} \cite{Einstein,Lusanna}.

This gauge symmetry also implies that the {\it material particle, string or
brane do not carry any local degrees of freedom in the presence of
dynamical gravity, described by the action functional, and 
not given as a fixed
background}
\footnote{\label{5}
Here we do not discuss matter fields which may `live' on the 
$p$--brane, like worldvolume vector gauge field of Dirichlet branes.}. 
Indeed, the local brane degrees of freedom could have 
a meaning by specifying its position in spacetime $M^D$, i.e. by relating 
a point/region of 
${\cal W}^{p+1}$ with a point/set of points in  $M^D$.
However, as in a general coordinate invariant theory the
spacetime point notion becomes `unphysical' [it is not a gauge invariant 
concept and
thus {\it cannot} be treated as an observable in so far as observables
are identified with gauge invariant entities], the only physically significant 
information is the existence of the brane worldvolume
and, if there are several branes, also the possible intersections
of their worldvolumes\footnote{This was the point of view 
accepted by Einstein 
\cite{Einstein}, see also \cite{Lusanna}, 
although only particles were considered that time.}. 
This implies, and it is implied by, 
the pure gauge nature of the {\it local} degrees of freedom of
a brane interacting with dynamical gravity.

The above discussion 
restores the `symmetry' between the bosonic and the fermionic degrees of 
freedom of superbrane. Indeed, in the gauge--fixed description of the 
supergravity---superbrane interaction, Eq. (\ref{SG+bb}), there are 
no traces of the fermionic degrees of freedom of the superbranes. 
Then the discussion above shows that
in this interacting systems the
superbrane does  not carry any {\it local}
bosonic degrees of freedom either. 
The global, topological degrees of freedom 
are of course still present [a closed brane differs from an open one
and, if the dynamical system involves two or more particles or branes,
intersecting branes are different from the nonintersecting ones]
and have no
fermionic counterparts. This, however, is natural in so far 
the topology of Grassmann algebra is trivial 
(`Ectoplasm has no topology',  J.S. Gates). 

\bigskip

{\bf 5. The Higgs effect in General Relativity interacting with material 
particles and 
extended objects}.

The pure gauge nature of the (super)brane coordinate
function in (super)gra\-vi\-ty--(super)brane interacting
system is actually not surprising. Indeed, it is well known (see 
\cite{Goldst}) that, for a brane in flat spacetime, the coordinate 
functions are essentially Goldstone fields corresponding to the 
generators of rigid translational symmetry broken by the presence of 
the $p$--brane.
When the brane interacts with {\it dynamical} gravity, the
rigid translational symmetry is replaced by the {\sl gauge}
diffeomorphism symmetry. Thus the coordinate functions
$\hat{x}^\mu(\tau)$ (and $\theta^{\check{\alpha}}(\tau)$) 
become the Goldstone  fields for this gauge symmetry. The Goldstone fields
for a gauge symmetry {\sl always} have a pure gauge nature.

On the other hand, the presence of the Goldstone fields also indicates that
the gauge diffeomorphism symmetry 
is spontaneously broken and that a counterpart of Higgs effect should occur
in the (super)gravity--(super-)$p$-brane interacting system.
This was discussed in detail in \cite{BdAIL3}.
The counterpart of the mass term for the gauge field
appears to be just the source term, {\it i.e.} the energy--momentum tensor
(\ref{Tmnp}) 
in the {\it r.h.s.} of the Einstein equation (\ref{EiEq0}).
However, as it was shown in \cite{BdAIL3}, this is {\sl not a mass term}.

When a field equation is considered in a curved spacetime, 
a term linear in the field entering this equation could be either mass term 
or a contribution from the nonvanishing curvature of the spacetime. 
The criterium which allows to define the notions 
of massive and massless (spin)tensor field in the curved  spacetime is  
the number of polarizations. A massless graviton should possess 
$D(D-3)/2$ physical polarizations ($2$ for $D=4$). 

Thus, the key point is whether in the (super)gravity--(super)brane 
interacting system the graviton keeps the same number of polarizations
as in the case of `free' gravity,  
even on the brane worldvolume $W^{p+1}$  dispite that 
 one uses a part of diffeomorphism invariance to fix the static gauge
(\ref{gaugeXp}). This is indeed the case \cite{BdAIL3} because,  
in accordance with the generalized 
Einstein--Grommer theorem \cite{EiGro,Gursey,BdAI},  
the Einstein equation (\ref{EiEq0}) with a singular energy--momentum 
tensor (\ref{Tmnp}) in the {\it r.h.s.} 
produces the $p$--brane equations of motion (\ref{pgeod}) as its 
selfconsistency conditions. 
In the static gauge  (\ref{gaugeXp}) this brane equations becomes
the conditions for the gravitation field on $W^{p+1}$ 
\cite{BdAIL3} and play there the role of the gauge fixing conditions
that were lost due to the loss of the diffeomorphism symmetry used
to fix (locally) the static gauge (\ref{gaugeXp}).

This can be understood from the perspective of the 
above discussion about absence of local degrees of freedom of a 
(super)brane in the presence of dynamical 
(super)gravity. Indeed, in the usual Higgs effect, 
the Goldstone {\it degrees of freedom} do exist and, after fixing the 
`unitary gauge' removing the Goldstone {\it fields}, their degrees of freedom 
reappear  as additional polarizations of 
the gauge field.  These make  the gauge field  massive, 
as also reflected by the appearance  of the mass term in the gauge field 
equations. In our case, as a brane does not carry any local degrees
of freedom in the presence of dynamical gravity, no degrees of
freedom are available to reappear as an additional polarization of graviton  
when the static gauge (\ref{gaugeXp}) is fixed. 
Hence the graviton keeps the same number of polarizations as in 
the absence of the brane and, thus, remains massless.

To conclude, let us note that, in general, 
when diffeomorphism invariance 
is spontaneously broken, a graviton might get a mass {\it provided} 
the  St\"uckelberg or Goldstone {\it degrees of freedom} are present and, 
thus, might appear as additional graviton polarizations.   
The studies of the massive graviton in AdS space 
\cite{PorratiDuffLiu,Porrati} refer just to the possibility 
of introducing the (vector) St\"uckelberg degrees of freedom in 
a selfconsistent manner. These  
might appear, {\it e.g.} as a bound 
state of two fields in a free CFT interacting with dynamical gravity 
on AdS space (but not on the Minkowski space) \cite{Porrati}. In contrast,
what has been  shown in \cite{BdAIL3} is  that  
 `material' branes, {\it i.e.} the branes described by a 
diffeomorphism invariant  action, 
although  provide Goldstone {\it fields} indicating the spontaneous 
breaking of the gauge diffeomorphism symmetry, 
these {\sl do not} carry  any {\it degrees of freedom} 
in the presence of dynamical gravity (see footnote \ref{5}).  
This is the reason why the graviton cannot acquire 
additional polarizations and cannot get mass in this case. 
It would be interesting to analyze in this perspective the 
`locally localized gravity' model of Ref. \cite{KR01}, 
where a region of $AdS_5$ space is restricted by two $AdS_4$ `hypersurfaces'.

\bigskip  

{\it Acknowledgments}. This work has been 
partially supported by the research grants BFM2002-03681,
from the Ministerio de Ciencia y Tecnolog\'{\i}a de Espa\~na 
and EU FEDER funds, $\# 383$ from Ucrainian FFR,  N 2000-254 from INTAS.

\bigskip

{\bf Appendix: Why the cosmological constant cannot be treated as a mass term.}

A good example to illustrate the above discussion on the relation of 
graviton  masslessness with diffeomorphism invariance 
is provided by the Einstein equations with a cosmological constant, 
${\cal G}_{\mu\nu}(g)= \Lambda g_{\mu\nu}$, whose vacuum solution is
given by anti--de Sitter ($AdS$) or by de Sitter ($dS$) space. 

Considering small excitations $h_{\mu\nu}$  
over the $AdS$  metric $g_{\mu\nu}^{AdS}$, $g_{\mu\nu}=
g_{\mu\nu}^{AdS} + h_{\mu\nu}$, one finds that the linearized Einstein
equation  ${\cal G}^{AdS}_{\mu\nu}(h)= \Lambda h_{\mu\nu}$ contains a
`mass--like' term, {\it i.e.} a term linear in field 
 $\Lambda h_{\mu\nu}$. However, this term might be not only a mass term but 
also a contribution form the nontrivial curvature of the AdS space.
This is indeed the case {\it as a model possessing full  
diffeomorphism invariance (and not involving St\"uckelberg degrees 
of freedom)  is considered}. 
  
To introduce the notion of a massless field 
in curved space one needs to have a smooth flat spacetime limit. 
This implies, in particular, that the definition of a massless 
field should produce the same number of polarizations in flat and curved 
spacetime. 

Now, as the linearized Einstein equations over AdS spacetime,  
${\cal G}^{AdS}_{\mu\nu}(h)= \Lambda h_{\mu\nu}$,  possess a linearized
diffeomorphism symmetry, one may restrict $h_{\mu\nu}$ by the same number
of gauge fixed conditions as in flat spacetime
(where one assumes $g_{\mu\nu}=
 \eta_{\mu\nu} + h_{\mu\nu}$ and considers the linearized equation   
${\cal G}^{\eta}_{\mu\nu}(h)=0$). This implies that the graviton 
in $AdS$ space has the same number of polarization as in the flat spacetime. 
Hence the graviton in $AdS$ space is massless and the cosmological 
constant cannot be treated as a mass term.    

One might also think of the case of small but nonvanishing cosmological 
constant $\Lambda$: $\Lambda \rightarrow 0$, but $\Lambda\not=0$ 
({\it cf.} the discussion in \cite{BdAIL3}), where 
one may consider decomposition over the flat spacetime, 
$g_{\mu\nu}=
 \eta_{\mu\nu} + h_{\mu\nu}$. Here if {\sl in the linear approximation} 
for the  weak field  $h_{\mu\nu}$ one were to find a term proportional to the 
field $h_{\mu\nu}$ with a constant coefficient, 
this could only be the mass term. 
The first impression might be that $\Lambda h_{\mu\nu}$, which appears 
in the decomposition of the Einstein equation 
${\cal G}_{\mu\nu}(\eta) + {\cal G}^{\eta}_{\mu\nu}(h) + {\cal O}(hh)
= \Lambda \eta_{\mu\nu} + \Lambda h_{\mu\nu}$ 
is just such a mass term. 
However, a more careful analysis shows that this is not the case 
({\it cf.} \cite{BdAIL3}). The flat metric does not solve the Einstein 
equation  with cosmological constant, 
${\cal G}_{\mu\nu}(g)= \Lambda g_{\mu\nu}$, but rather solves  `free' 
Einstein equation, ${\cal G}_{\mu\nu}(\eta)=0$. As a result, 
the decomposition of the Eisntein equation reads
\begin{equation}\label{linEi}
{\cal G}^{\eta}_{\mu\nu}(h) + {\cal O}(hh)
= \Lambda \eta_{\mu\nu} + \Lambda h_{\mu\nu}\; .
\end{equation}
Before turning to the linear approximation in the weak field $h$, which is 
needed to search for possible mass term, one has to be convinced that 
the zero--order approximation is selfconsistent. 
The {\it l.h.s.} 
of Eq. (\ref{linEi}) does not contain a zero--order term at all. 
However, if one considers $\Lambda h_{\mu\nu}$ 
as a first order term in the weak field $h_{\mu\nu}$, then 
one has to conclude that $\Lambda \eta_{\mu\nu}$ is the term of zero order; 
in this setup the zero order approximation for the equation (\ref{linEi}) 
reads $0= \Lambda \eta_{\mu\nu}$ and implies vanishing cosmological constant
$\Lambda=0$. 
Thus, to consider the case of nonvanishing but very small 
cosmological constant, and, at the same time, 
to make a selfconsistent weak field approximation {\it over a 
flat background},  
the only posibility is to assume that 
$\Lambda$ is of the same order of `smallness' as $h_{\mu\nu}(x)$, 
$\Lambda \sim h_{\mu\nu}(x)$. [Actually this is natural from the prespective 
of decomposition over the AdS background, $g_{\mu\nu}= g_{\mu\nu}^{AdS} + 
h^0_{\mu\nu}$, as for $\Lambda\rightarrow 0$ $g_{\mu\nu}^{AdS}= 
\eta_{\mu\nu}+h_{\mu\nu}^{AdS}$ with a small $h_{\mu\nu}^{AdS}$; the above 
$h_{\mu\nu}$ is given by $h_{\mu\nu}=h_{\mu\nu}^{AdS}+h_{\mu\nu}^{0}$].
In this setup, however, $\Lambda h(x)\sim h(x)h(x)$. 
 Hence, the term $\Lambda h_{\mu\nu}$ can be considered only together with 
${\cal O}(hh)$ terms in the {\it l.h.s.} of Eq. (\ref{linEi}); it is a term 
of second order in the weak field approximation. The first order approximation 
is given by ${\cal G}^{\eta}_{\mu\nu}(h)
= \Lambda \eta_{\mu\nu}$ which contains the constant source term rather than
a term linear in the weak field. 
Thus no mass term appears in the selfconsistent linear approximation.
[The constant source does not change the
number of field theoretical degrees of freedom; see {\it e.g.} \cite{BdAI}]. 

This shows in a different way the masslessness of the graviton in
$AdS$ space, {\it i.e.}, that the cosmological constant cannot be 
considered as a mass term.

\end{document}